\documentclass{aa} 
\usepackage{natbib}
\usepackage{graphicx}
\usepackage{epsfig}
\usepackage{times}
\usepackage{rotating}
\usepackage{float}
\usepackage{setspace}
\usepackage{lscape}
\usepackage{epstopdf}
\usepackage{tabularx}
\usepackage{graphicx} 
\usepackage[caption = false]{subfig}
\usepackage[usenames]{color}
\usepackage[varg]{txfonts}

\begin{document}

\title{Fossil group origins}

\subtitle{X. Velocity segregation in fossil systems}

\authorrunning{S. Zarattini et al.}

\titlerunning{Velocity segregation in fossil systems}

\author{S. Zarattini\inst{1,2}, J. A. L. Aguerri\inst{3,4}, A. Biviano\inst{5}, M. Girardi\inst{5,6}, E. M. Corsini\inst{7,8}, and E. D'Onghia\inst{9,10}}

\institute{IRFU, CEA, Universit\'e Paris-Saclay, F-91191 Gif-sur-Yvette, France\\
	 \email{stefano.zarattini@cea.fr}
\and AIM, CEA, CNRS, Universit\'e Paris-Saclay, Universit\'e Paris Diderot, Sorbonne Paris Cit\'e, F-91191 Gif-sur-Yvette, France
\and Instituto de Astrof\'isica de Canarias, calle V\'ia L\'actea s/n, E-38205 La Laguna, Tenerife, Spain
\and Departamento de Astrof\'isica, Universidad de La Laguna, Avenida Astrof\'isico Francisco S\'anchez s/n, E-38206 La Laguna, Spain
\and INAF-Osservatorio Astronomico di Trieste, via Tiepolo 11, I-34143 Trieste, Italy
\and Dipartimento di Fisica, Universit\`{a} degli Studi di Trieste, via Tiepolo 11, I-34143 Trieste, Italy
\and Dipartimento di Fisica e Astronomia ``G. Galilei'', Universit\`{a} di Padova, vicolo dell'Osservatorio 3, I-35122 Padova, Italy
\and INAF-Osservatorio Astronomico di Padova, vicolo dell'Osservatorio 2, I-35122 Padova, Italy
\and Astronomy Department, University of Wisconsin, 475 N Charter Street, Madison, WI, 53706
\and Harvard Center for Astrophysics, 60 garden Str., 02138, Cambridge, MA}

\date{\today}

\abstract{}
{We want to study how the velocity segregation and the radial profile of the velocity dispersion depend on the prominence of the brightest cluster galaxies (BCGs).}
{We divide a sample of 102 clusters and groups of galaxies into four bins of magnitude gap between the two brightest cluster members. We then compute the velocity segregation in bins of absolute and relative magnitudes. Moreover, for each bin of magnitude gap we compute the radial profile of the velocity dispersion.}
{When using absolute magnitudes, the segregation in velocity is limited to the two brightest bins and no significant difference is found for different magnitude gaps. However, when we use relative magnitudes, a trend appears in the brightest bin: the larger the magnitude gap, the larger the velocity segregation. We also show that this trend is mainly due to the presence, in the brightest bin, of satellite galaxies in systems with small magnitude gaps: in fact, if we study separately central galaxies and satellites, this trend is mitigated and central galaxies are more segregated than satellites for any magnitude gap. A similar result is found in the radial velocity dispersion profiles: a trend is visible in central regions (where the BCGs dominate) but, if we analyse the profile using satellites alone, the trend disappears. In the latter case, the shape of the velocity dispersion profile in the centre of systems with different magnitude gaps show three types of behaviours: systems with the smallest magnitude gaps have an almost flat profile from the centre to the external regions; systems with the largest magnitude gaps show a monothonical growth from the low values of the central part to the flat ones in the external regions; finally, systems with $1.0 < \Delta m_{12} \le 1.5$ show a profile that peaks in the centres and then decreases towards the external regions.}
{We suggest that two mechanisms could be responsible for the observed differences in the velocity segregation of the BCGs: an earlier formation of systems with larger magnitude gap or a more centrally-concentrated halo. However, the radial profiles of the  velocity dispersion confirm that central galaxies are more relaxed, but that the satellite galaxies do not seem to be affected by the magnitude gap.}


\maketitle

\section{Introduction}
It is well known that the galaxy population is different in the field and in clusters. Many authors \citep[][amongst others]{Melnick1977,Whitmore1993} showed that red early-type galaxies are located in the central (denser) regions of nearby galaxy clusters, whereas blue star-forming late-type galaxies are found at larger radii, where the density is lower. \citet{Dressler1980} showed that there is a correlation between the fraction of galaxies of different morphological types and the local projected galaxy density. He then concluded that the presence of different morphological types in different regions of the clusters is not dependent on the global conditions related to the cluster's environment, but it is more connected with the local clustering. However, \citet{Sanroma1990} showed that this morphological trend is driven by the projected radius and not by the surface density.

A phenomenon that is related to this spatial segregation of galaxies is their segregation in the velocity space. The study of this latter segregation requires a bigger observational effort, since deep spectroscopy is needed. However, various studies were conducted on this topic and significant differences in the velocity distribution are found for different galaxy populations. As an example, a difference in velocity dispersion measures using red and passive galaxies or blue star-forming galaxies in clusters was found by many authors \citep[][]{Moss1977,Sodre1989,Biviano1992,Biviano1996,Biviano1997,Scodeggio1995,Goto2005,Sanchez-Janssen2008}. Another type of known segregation is that of massive and luminous galaxies, that are found to be segregated in velocity with respect to smaller and fainter galaxies. This effect was reported for the first time by \citet{Rood1968} and then further analysed by other authors \citep[e.g.][and references therein]{Biviano1992}. These authors showed that only the most luminous galaxies are segregated and that brighter galaxies have lower velocities. The result was confirmed more recently by \citet{Adami1998}, \citet{Girardi2003}, \citet{Goto2005}, and \citet{Ribeiro2013} and it is generally explained by invoking physical processes that are able to transfer the kinetic energy of galaxies mainly to the dark matter (DM) particles. The process responsible of this effect is usually identified with dynamical friction \citep{Chandrasekhar1943,Sarazin1986}: a massive galaxy that is falling into a cluster interacts with other, smaller, galaxies. The smaller galaxies are gaining energy and momentum in the interaction at the expense of the massive galaxy. Moreover, the falling galaxy is also suffering a dynamical friction from the DM halo of the cluster. On the contrary, the process of violent relaxation is expected to produce a velocity distribution that is independent of galaxy mass \citep{Lynden-Bell1967}.

The existence of velocity segregation in galaxy clusters was also considered as a sign of dynamical evolution \citep{Coziol2009}. In fact, dynamical friction requires a long time to produce its effect on the more massive galaxies. However, \citet{Skibba2011} showed that between 25\% and 40\% of the brightest galaxies are satellites instead of central. We already showed \citep[][hereafter Paper VII]{Zarattini2016} that the majority of the BCGs located in fossil groups (FGs) seems to lie at the centre of the potential well, not showing a significant peculiar velocity, a result recently confirmed by \citet{Gozaliasl2019}. This could be interpreted as velocity segregation, but to really address this issue we also have to study the peculiar velocities of satellite galaxies. In addition, as in \citet[][hereafter Paper V]{Zarattini2015}, we now have enough statistics to understand if there is a dependence between the magnitude gap and the velocity segregation of galaxies.

Moreover, we also focus our attention on how the radial velocity dispersion profiles vary with the magnitude gap. These profiles, together with the projected number density and mass profiles, are able to give hints on the type of orbits that dominate a cluster \citep{Biviano2004,Aguerri2017}. In the context of this work, we are interested in studying if different orbits are found in systems with different magnitude gaps. \citet{Biviano2004} showed that the velocity dispersion profiles can be distinct for different galaxy populations: early-type galaxies show a decreasing radial profile towards the centre of the clusters, whereas late-type galaxies show a clear increase in the centre. This is supposed to be connected with different types of orbits: isotropic for early-type galaxies and more radial for late-type ones. Recently, \citet{Aguerri2017} also confirmed that there is a segregation of orbits depending on the luminosity of galaxies: more luminous galaxies are in less radial orbits than fainter galaxies. A similar orbital difference for galaxies of different mass was found by \citet{Annunziatella2016}, but in only the inner regions of clusters.

\defcitealias{Aguerri2011}{Paper~I}

This work is part of the Fossil Group Origins (FOGO) project, a program presented in \citet{Aguerri2011} in which we studied different aspects of FGs, a particular type of galaxy aggregation dominated at optical wavelengths by a massive and luminous central galaxy, at least two magnitudes brighter than the second brightest member in the $r$-band ($\Delta m_{12} \ge 2$). The detailed study of the sample was presented in \citet[][hereafter Paper IV]{Zarattini2014}. The project studied the properties of the BCGs in FGs \citep{Mendez-Abreu2012}, the X-ray versus optical scaling relations \citep{Girardi2014}, the global X-ray scaling relations \citep[][hereafter Paper VI]{Kundert2017}, the dependence of the luminosity function on the magnitude gap (Paper V), the presence of substructures in FGs (Paper VII), and the stellar populations in FG BCGs \citep{Corsini2018}. Moreover, we presented a case of transitional fossil system in \citet{Aguerri2018}.

The paper is structured as follows: in Sect. \ref{sample} we present the sample used for this work, in Sect. \ref{results} we present the results on both the velocity segregation and the radial velocity dispersion profiles, in Sect. \ref{discussion} we discuss the results, and in Sect. \ref{conclusions} we draw our conclusions.

The cosmology adopted in this paper, as in the rest of the FOGO papers, is $H_0 = 70$ km s$^{-1}$ Mpc$^{-1}$, $\Omega_\Lambda=0.7$, and $\Omega_{\rm M}$=0.3.

\section{Sample}
\label{sample}
We build our sample by combining two different datasets, as done for the study of the dependence of the luminosity function on the magnitude gap (Paper V). In the first sample (hereafter S1) we have the 34 FG candidates proposed by \citet{Santos2007} and analysed by the FOGO team in Paper IV. The spectroscopic completeness of the S1 sample is more than 70\% down to $m_r=17$ and more than 50\% down to $m_r=18$. For the S1 sample we have a total of 1244 available velocities (26 clusters), of which 579 turned out to be members of the respective cluster. We refer the reader to Paper IV for detailed information of the S1 sample. We were able to confirm that 15 out of 34 are genuine FGs, 7 are non-FGs, and the other 12  remain to be confirmed. We defined as ``genuine fossil'' a group or cluster of galaxies that accomplishes one of the two definitions that follow: i) it has a magnitude gap of at least 2 magnitudes between its two brightest member galaxies ($\Delta m_{12} \ge 2$) in the $r$ band within half the virial radius or ii) it has a magnitude gap of at least 2.5 magnitudes between the first and the fourth brightest members ($\Delta m_{14} \ge 2.5$) in the $r$ band within half the virial radius (see Paper IV for details). For this work we use only systems with $z \le 0.25$. This cut in redshift allows us to reach the dwarf regime in all the clusters. After its application the number of clusters in the S1 sample reduces to 26. In this sample, the BCGs are the starting point for the selection of the sample and are taken from the Sloan Digital Sky Survey {\it luminous red galaxies} sample. Details on the selection criteria can be found in \citet{Santos2007}. The median mass of clusters in the S1 sample, computed from the line-of-sight velocity dispersion using Eq. 1 from \citet{Munari2013}, is log M$_{200}= 14.21\pm0.42$ dex.

The S1 sample is biased towards systems with large magnitude gaps, since it was selected to find new FGs. In particular, the mean value of the $\Delta m_{12}$ parameter is $\Delta m_{12}\sim 1.5$ and only 4 systems have $\Delta m_{12} \le 0.5$. We, thus, add a second sample (hereafter S2) taken from \citet{Aguerri2007}. These systems were selected as clusters with $z \le 0.1$ from the catalogs of \citet{Zwicky1961}, \citet{Abell1989}, \citet{Voges1999}, and \citet{Bohringer2000}. They were observed in the Sloan Digital Sky Survey Data Release 4 \citep[SDSS-DR4,][]{Adelman-McCarthy2006}. The S2 sample consists of 88 clusters, but we only used those whose $\Delta m_{12}$ was spectroscopically confirmed. This criterium reduced the number of systems in the S2 sample to 76, with a mean $\Delta m_{12} \sim 0.7$. The spectroscopic completeness of the S2 sample is more than 85\% down to $m_r=17$ and more than 60\% down to $m_r=18$. In the S2 sample, there are a total of 5977 velocities available, of which 3886 turned out to be members. In S2, the centre of the cluster was determined using the peak of the X-ray emission (when available) or the peak of the galaxy distribution. Then, the BCG was selected to be the brightest galaxy of the central region. The mean difference between the centre of the cluster and the selected BCG is 150 kpc \citep[see][for details]{Aguerri2007}. The median mass of clusters in the S2 sample, computed as for S1, is log M$_{200}= 14.27\pm0.37$ dex. 
A detailed comparison between S1 and S2 was presented in Paper V and references therein. The number of clusters obtained by combining S1+S2 and applying the described cut in redshift is thus 102. We stress that these are exactly the same systems used in Paper V and that there is no intersection between the two samples.

\section{Results}
\label{results}
In this section we present the results of the study of the dependence of the velocity segregation (\ref{velocity}) and of the velocity dispersion profile (\ref{dispersion}) on the magnitude gap.

\begin{figure}[t]
\includegraphics[width=0.5\textwidth]{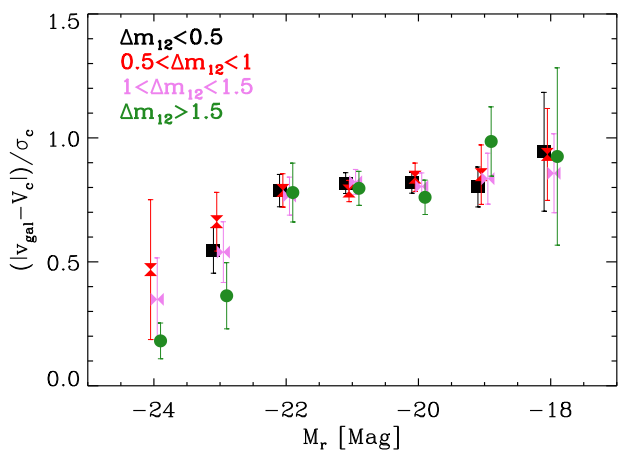}
\caption{Dependence of the velocity segregation on luminosity in bins of the magnitude gap for galaxies within R$_{200}$. We show with black filled squares those clusters with $\Delta m_{12} \le 0.5$, with red filled hourglasses those with $0.5 < \Delta m_{12} \le 1.0$, with violet filled bowties those with $1.0 < \Delta m_{12} \le 1.5$, and with green filled circles those systems with $\Delta m_{12} \ge 1.5$. The error bars uncertainties of the mean, computed as the standard deviation of the normalised velocity divided by the square root of the total number of galaxies in each absolute-magnitude bin.}
\label{vel_seg_abs1}
\end{figure}

\subsection{Dependence of the velocity segregation on the magnitude gap}
\label{velocity}

Following what we did in Paper V, we divide our sample of 102 clusters and groups in 4 subsamples of different $\Delta m_{12}$. In particular, the first subsample includes all clusters with $\Delta m_{12} \le 0.5$, the second one those with $0.5 < \Delta m_{12} \le 1.0$, the third subsample those with $1.0 < \Delta m_{12} \le 1.5$, and finally the last subsample has all the systems with $\Delta m_{12} > 1.5$. According to this division, we have 31, 24, 26, and 21 systems in the first, second, third, and fourth subsample, respectively. The values of the magnitude gaps used to divide the sample are arbitrary and chosen to have more that 20 systems per bin, in order to obtain robust statistical results. The median velocity dispersion of the four different subsamples are $557\pm171$ km s$^{-1}$, $617\pm159$ km s$^{-1}$, $587\pm200$ km s$^{-1}$, and $545\pm206$ km s$^{-1}$, respectively. The corresponding median M$_{200}$ masses (given in logarithmic units) are $14.27\pm0.38$ dex, $ 14.40\pm0.35$ dex, $14.33\pm0.40$ dex, and $14.21\pm0.48$ dex.

\begin{figure*}
\includegraphics[width=0.99\textwidth]{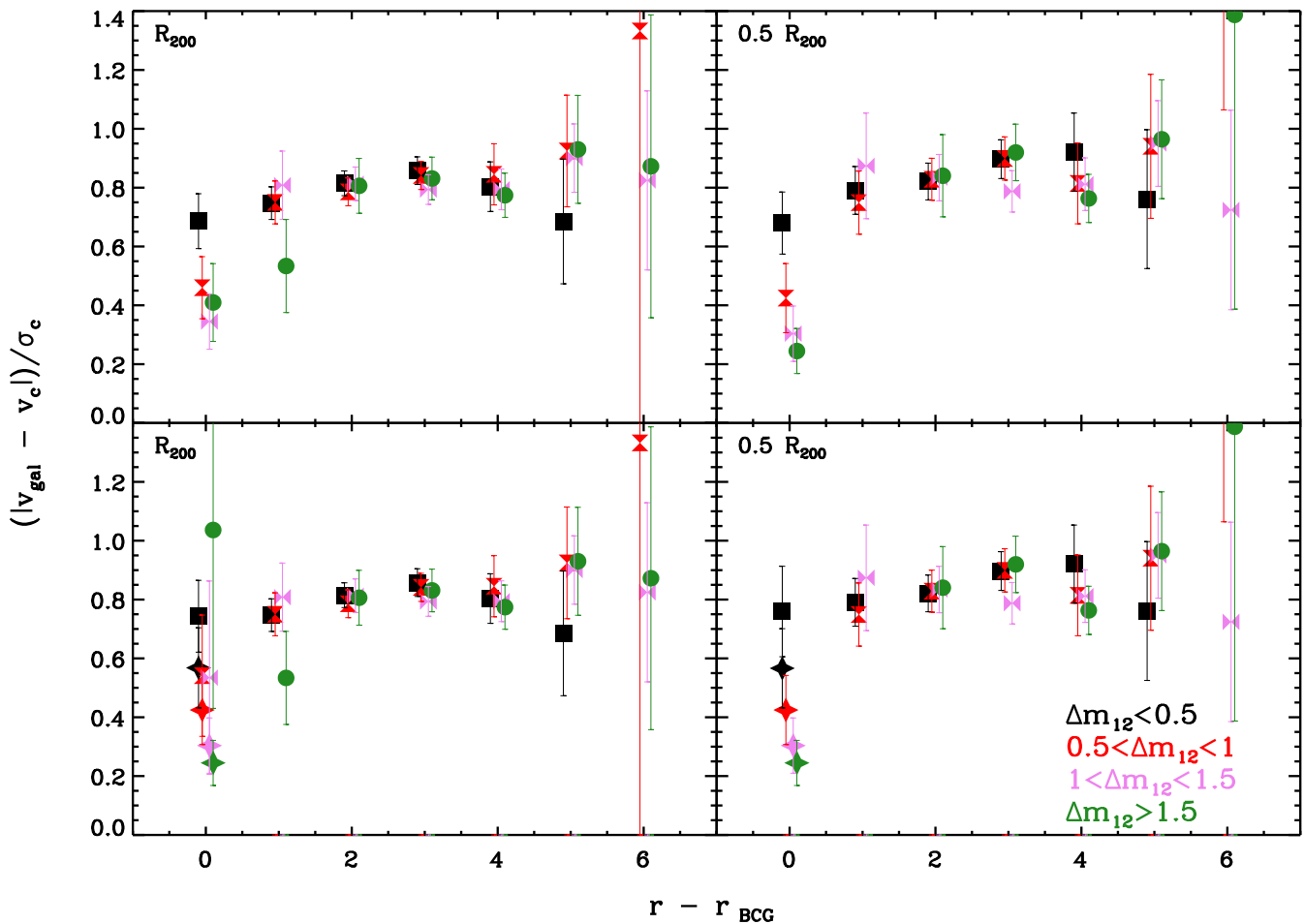}
\caption{Dependence of the velocity segregation with the magnitude gap obtained using relative magnitudes. In the top left panel we show the result obtained with all the galaxies within 1 R$_{200}$, whereas in the top right panel we use only galaxies within 0.5R$_{200}$. In the lower panels we show the same results, but separating central galaxies (filled stars) from satellites. The remaining symbols and color code are the same as in Fig. \ref{vel_seg_abs1}.}\label{vel_seg_rel}
\end{figure*}

For each member galaxy, we have a velocity that comes from our own spectroscopy or SDSS and magnitudes from SDSS-DR7. In particular, throughout the paper we use the model magnitude in the $r$ band and all absolute magnitudes are K$-$corrected following \citet{Chilingarian2010}. 

For each cluster of the S1 sample, R$_{200}$ was obtained from X-ray data and members were selected using a two-step procedure applied to the galaxies in the region within R$_{200}$. First, we used DEDICA \citep{Pisani1993,Pisani1996}, which is an adaptive kernel procedure that works under the assumption that a cluster corresponds to a local maximum in the density of galaxies. Then, we adopted the likelihood ratio test \citep{Materne1979} to assign a membership probability to each single galaxy relative to an identified cluster. The details on these procedures are described in Paper IV.
For the clusters in the S2 samples, \citet{Aguerri2007} also adopted a two-steps procedure in which the first step was the gapping procedure proposed by \citet{Zabludoff1990}. They also applied the KMM algorithm \citep{Ashman1994} to estimate the statistical significance of bi-modality in the main peak identified in the first step. Once the members of the clusters are identified, they are used to compute the value of R$_{200}$ for each cluster. Details on this procedure can be found in \citet{Aguerri2007}. 
The two procedures applied to the S1 and S2 samples are robust to interlopers, thus granting a reliable measure of the cluster global properties. Once R$_{200}$ and member galaxies were known, we compute the distance of each galaxy from the centre of the cluster (defined as the BCG position), which will be used to compute the velocity dispersion profile in Sect. \ref{dispersion}.

The velocity segregation is computed in bins of absolute magnitude. For each galaxy we compute its normalised velocity as 
\begin{equation}
v_{\rm gal}^{\rm norm}=\frac{\left| v_{{\rm gal}}-v_{\rm c} \right|}{\sigma_{\rm c}}
\end{equation}
where $v_{\rm gal}$ is the recessional velocity of the galaxy itself, whereas $v_{\rm c}$ and $\sigma_{\rm c}$ are the mean velocity and the velocity dispersion of the corresponding cluster, computed within at least 0.5 R$_{200}$ and after removing velocity interlopers \citep[see][for details]{Zarattini2014}. Then, for each absolute-magnitude bin we compute the mean value of the normalised velocity for all the clusters that have at least a galaxy with the required magnitude.

We present in Fig. \ref{vel_seg_abs1} the dependence of the velocity segregation on magnitude gap. The plot is computed using all galaxies within the virial radius. It can be seen that the segregation in velocity appears only in the two most-luminous magnitude bins. No significant trend appears with the magnitude gap, although systems with the larger gaps seems to show a larger segregation. It is worth noticing that clusters with $\Delta m_{12} \le 0.5$ do not have galaxies in the brightest magnitude bin (e.g. no galaxies with $-24.5 \le r \le -23.5$). The Spearman rank correlation test confirms that a correlation is present in each of the four subsamples. Moreover, in order to assess the significance of the segregation, we shuffle 100 times the magnitude gaps among the 4 subsamples and recompute $v_{\rm norm} (M_r > -23) - v_{\rm norm} (M_r < -23)$ for each case. We find that only in 10\% of the cases the velocity segregation is as high as the one found for the $\Delta m_{12} > 1.5$ subsample.

As we did in Paper V, we compute also the \emph{relative} velocity segregation by subtracting the magnitude of the central galaxy to all the magnitudes. As a result, all the BCGs are located in the same bin, independently of their magnitude and on the mass of their hosting group/cluster. This is useful when comparing objects with different masses as clusters and groups in order to highlight differences that originate directly from the magnitude gap. In fact, as we showed in Paper V, the central galaxies of groups are fainter than those of clusters and they can lie in a region where there are a lot of intermediate-mass galaxies in clusters. For this reason, in Fig. \ref{vel_seg_abs1} the impact of the magnitude gap can be mitigated.

 We present the relative velocity segregation in Fig. \ref{vel_seg_rel}. In the top left panel, we show the result obtained within $R_{200}$: the color code is the same as in Fig. \ref{vel_seg_abs1} and we can see that the segregation in velocity seems to be limited to galaxies that are as bright as the BCGs. Moreover, it seems that there can be a trend in the first magnitude bin: here galaxies located in clusters with larger magnitude gaps seem to be more segregated. However, the trend is not strong. We then compute the same quantities within 0.5 $R_{200}$ and we show the results in the top right panel of Fig. \ref{vel_seg_rel}. Now the trend with magnitude gap is stronger because the magnitude gap is computed for galaxies within 0.5R200, while in the left panels we include also galaxies out to $R_{200}$. Thus, it is possible that other galaxies as bright as the BCGs can be found at distances larger that 0.5 $R_{200}$, not affecting the magnitude gap but affecting the observed segregation. The trend shows that the larger the gap, the more segregated are the (central) galaxies in velocity. As we found for the LFs, there is a statistical difference larger that 3$\sigma$ between the two most-extreme cases (namely, $\Delta m_{12} \le 0.5$ and $\Delta m_{12} > 1.5$), with the other two magnitude-gap bins located in the middle and following the trend set by the magnitude gap itself.
 
 Again, as we did for Fig. \ref{vel_seg_abs1}, we compute the difference in velocity between the two brightest bins for the subsample of clusters with $\Delta$ $m_{12} \ge1.5$. This difference is the largest that we found (in Fig. 2, this is the difference between the two left-most green points in the top-right panel) and it is measured to be 0.6. We then shuffle 100 times the magnitude gap of each cluster and each time we compute again the resulting relative velocity segregation for the 4 subsamples (that have no relation with the magnitude gap at this point). Thus, for each subsample we compute the velocity segregation for the first two bins and compare it with our reference value, finding that the new values are never larger than 0.6.
 
 We also show in the two lower panels of Fig. \ref{vel_seg_rel} the same results, but splitting the galaxy population of each cluster in two components: the BCGs and the satellite galaxies. The two most interesting results in these panels are i) the trend in the first magnitude bin is still visible, although it is weaker and limited to BCGs alone, and ii) satellites do not seem to suffer from velocity segregation, independently of their magnitude. In fact, the bottom right panel of Fig. \ref{vel_seg_rel} clearly shows that satellite galaxies located in the two brightest magnitude bins have the same normalised velocity as satellite galaxies in the fainter magnitude bins. On the other hand, when comparing the top and bottom panels it can be seen that the velocity segregation in the brightest magnitude bin is stronger when using only BCGs (bottom panels) than when also satellites are included. This is particularly evident for the first subsample, where massive satellites can be found in the brightest bin. However, a trend in velocity segregation is visible also in the bottom panels, but it is mitigated, especially within 0.5$R_{200}$.
\subsection{Dependence of the velocity dispersion profile on the magnitude gap}
\label{dispersion}

We also study the cumulative and differential velocity dispersion profiles, $\sigma^{(n)}_{{\rm cum}} (<R)$ and $\sigma^{(n)}_{{\rm diff}} (R)$ respectively, in the different magnitude-gap bins. The differential quantities are computed at $R=0.025$ R$_{200}$, $R=0.05$ R$_{200}$, $R=0.1$ R$_{200}$, $R=0.2$ R$_{200}$, $R=0.5$ R$_{200}$, $R=1$ R$_{200}$. The cited quantities and the 68\% uncertainties are computed using the biweight estimator of the ROSTAT package \citep{Beers1990}.

\begin{figure}[t]
 \includegraphics[width=0.5\textwidth]{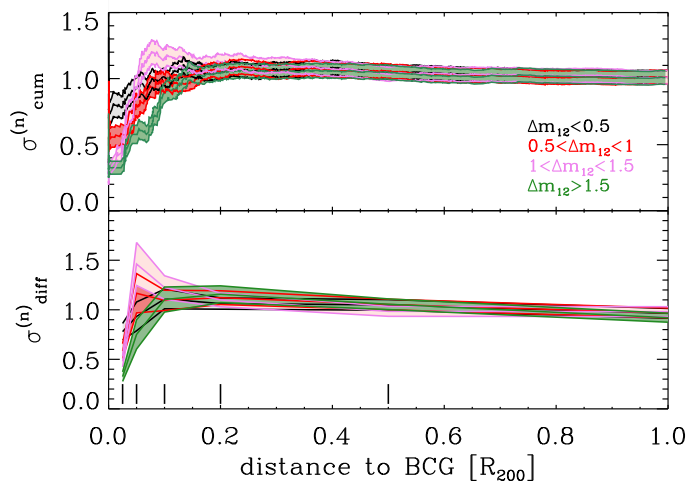}
 \caption{Cumulative (upper panel) and differential (lower panel) velocity dispersion profiles for clusters in different magnitude-gap bins. The color code is the same as in Fig. \ref{vel_seg_abs1} and shaded areas represent 1$\sigma$ uncertainties. The small vertical lines represent the radii at which the differential quantities are computed.}
 \label{vel_disp}
\end{figure}

In Fig. \ref{vel_disp} we present the results. In the very central part a trend seems to appear: the clusters with the largest $\Delta m_{12}$ have a lower value of velocity dispersion in the centre. However, the trend is not as clear as it was for the velocity segregation, because clusters with $1.0 < \Delta m_{12} \le1.5$ have a small velocity dispersion in the centre, which roughly triples at 0.07 $R_{200}$. A difference remains in the two most-extreme cases. In fact, it can be seen that at large radii all the velocity dispersion profiles become flatter at a value that is close to unity, which is the value of the velocity dispersion computed using all the galaxies in the sample. However, the way in which the different subsamples selected using the magnitude gap reach this limit is different: clusters with $\Delta m_{12} \le 0.5$ reach it faster and the ``jump'' between the most central values and the plateau is smaller than that of clusters with $\Delta m_{12} > 1.5$. It is worth noticing that the first (left-most) points of Fig. \ref{vel_disp} confirms the results presented in Fig. \ref{vel_seg_rel} for the BCGs.

\begin{figure}
\includegraphics[width=0.5\textwidth]{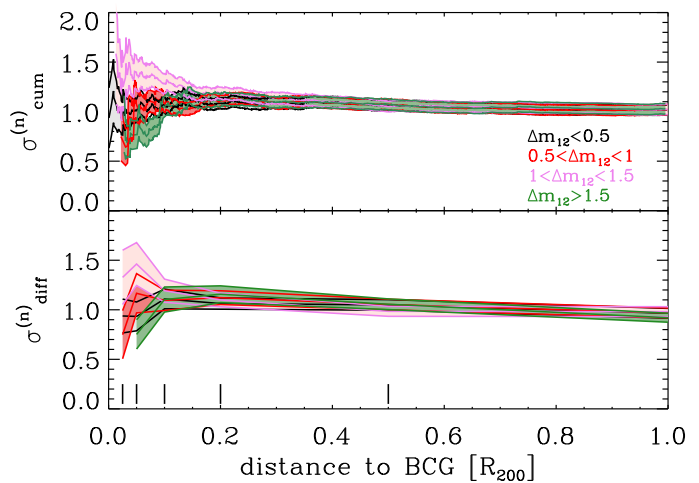}
\caption{Same as Fig. \ref{vel_disp}, but restricted to satellite galaxies.}
\label{vel_disp_sat}
\end{figure}

In Fig. \ref{vel_disp_sat} we show the same analysis, but for satellite galaxies only. Here the trend disappears and we can identify three different behaviours: systems in which also the satellite galaxies have a small velocity dispersion in the centre of the clusters (clusters with $0.5 < \Delta m_{12} \le 1.0$ and $\Delta m_{12} > 1.5$), systems with an almost flat profile ($\Delta m_{12} \le 0.5$), and systems with a higher velocity dispersion in the centre that in the more external parts (clusters with $1 < \Delta m_{12} \le 1.5$)

\section{Discussion}
\label{discussion}

We show in Sect. \ref{vel_seg_rel} that there is a dependence of the velocity segregation on the magnitude gap. In particular, there is a trend for systems with larger magnitude gaps to have more-segregated bright galaxies. Moreover, the segregation is clearly limited to the central galaxies alone. These two results suggest that there is some connection between the magnitude gap and the consequences it has on the evolution of the BCGs, but that this connection is not affecting satellite galaxies. A possible explanation is that the DM halos in systems with larger magnitude gaps are found to be more centrally concentrated \citep{DOnghia2005,vonBenda2008,Ragagnin2018}. Those halos can thus create a deeper potential well, able to enhance the dynamical friction with the central galaxy. In fact, it is known \citep[e.g.][]{Adhikari2016} that a DM subhalo orbiting inside a larger halo will experience dynamical friction due to the density of DM particles in the host halo, with a rate
\begin{equation}
\frac{d{\mathbf v}}{dt} \propto -\frac{G^2 M \rho}{v^3} {\mathbf v} f({\rm v} | \sigma).
\end{equation}
In this equation, $M$ is the mass of the subhalo, ${\mathbf v}$ is its relative velocity with respect to the host, $\rho$ is the local density of the host halo, and $f({\rm v} | \sigma)$ is the fraction of field particles with velocity less than $\lvert v \rvert$. The result is that a massive halo going through a more centrally-concentrated host should experience a stronger deceleration and it would end at the bottom of the potential well in a shorter time than the same halo in a sparser host \citep[see also][]{Jiang2008}.

A similar argument can be used if systems with large magnitude gaps form at an earlier epoch than systems with smaller magnitude gaps. In fact, in this case the former would have more time to slow down the central galaxies, without the need to assume more centrally-concentrated DM halos. This early formation was predicted using hydrodynamical simulations by \citet{DOnghia2005} and \citet{vonBenda2008}, but more recent simulations question this point. In particular, in Paper VI we showed, using the Illustris cosmological hydrodynamical simulation, that the early mass assembly history of fossil systems (e.g. systems with $\Delta m_{12} \ge 2$) is the same as for non-fossil systems. Differences rise at $z < 0.4$, when fossil systems stop accreting galaxies and have enough time to merge the big satellites into their BCG, thus creating the gap. In particular, using semi-analytical modelling, \citet{Farhang2017} noticed that FGs evolve faster then non-FGs in the redshift range $0.4>z>0.1$ and they than evolve as non-FGs since $z=0.1$. On the other hand, it is worth noticing that our results seem to confirm that systems with small gaps could be cluster mergers as suggested, for example, by \citet{Trevisan2017}.

Recently, \citet{Barsanti2016} have studied a sample of 41 clusters in the range $0.4 \le z \le 1.5$ and have looked for luminosity segregation in velocity space. They found evidence of segregation for all those galaxies brighter than the third most luminous galaxy in each cluster. Moreover, the more luminous is the galaxy, the lower is its velocity. This result is also in agreement with that reported for a local sample by \citet{Biviano1992}. Our systems are divided in magnitude gap bins and the growing of the gap can also affects velocity segregation: in fact, the larger the gap, the more luminous (and massive) the BCG and, as a consequence, the more segregated the BCG should be in velocity. However, our findings show that only the BCGs are strongly segregated in velocity. No hints are found that bright and massive satellites suffer velocity segregation.

Unfortunately, the result of the dependence of the velocity segregation on the magnitude gap is not able to favor one of the two scenarios proposed in the last paragraph with respect to the other. Thus, we also study the radial velocity dispersion profile in order to obtain more information and try to favor one of the previous scenarios on the other. From a theoretical point of view, the computation of the ``inverted'' Jeans equation, that is used to compute the velocity-anisotropy profile \citep[see][ for details]{Biviano2004} depends on the number density of galaxies, the mass and velocity dispersion profiles. Thus, assuming the same mass profile for all the clusters, the observed differences in the velocity dispersion profiles can depend on the spatial distribution, on different orbits of galaxies or on different dark matter concentration. 

\begin{figure}
\includegraphics[width=0.45\textwidth]{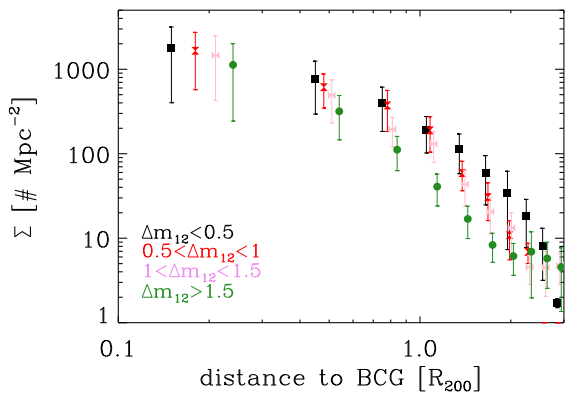}
\caption{Surface density profiles of galaxies in the different subsamples. The color code is the same as in Fig. 1.}
\label{spatial}
\end{figure}

We then compute the spatial distribution of galaxies in the four subsamples and present the result in Fig. \ref{spatial}. There seems to be a small trend in which objects with larger $\Delta m_{12}$ have steeper galaxy densities. However, the individual points for the different subsamples are all in agreement within the uncertainties and we also perform a linear fit to the spatial distribution of the different subsamples, finding no differences in the slopes at $1\sigma$ level. We also calculate the cluster number concentration for our four subsamples. To do so, we firstly compute the radial completeness profiles for each cluster separately. These were obtained by selecting all galaxies with and without redshift in the chosen magnitude range (the upper quartile of the magnitude distribution of members), binning in radii the two, and making the number ratio in each bin. We then assign to each member galaxy the completeness value that corresponds to the radial bin that includes the galaxy itself.
Then, we stack the clusters according to their $\Delta m_{12}$ and we fit a projected Navarro-Frenk-White profile \citep[NFW,][]{Navarro1997} to the stacked profiles, taking the radial completeness correction into account. The fits are done out to $R_{200}$, since at larger radii not all the clusters are contributing. The concentrations that we find are $c_1=2.0\pm0.1$,  $c_2=2.1\pm0.2$, $c_3=1.9\pm0.2$, and $c_4=2.5\pm0.4$ for the four subsamples of different $\Delta m_{12}$. These values are quite low compared to theoretical predictions for the concentration of DM, also considering that \citet{Biviano2017} found that galaxies and DM concentrations are similar for nearby clusters. On the other hand, the values we find are quite in agreement with other observational works, like \citet{Lin2004} ($c=2.9 \pm 0.2$), \citet{vanderBurg2015}  ($c=2.03 \pm 0.2$), and references therein.
Although the concentrations we found for our four subsamples are compatible with one another, it is worth noticing that systems with larger mass seem to have higher concentrations. This result was already found in \citet{Farhang2017}: in particular, in their fig. 8, they showed that systems with large magnitude gap have systematically a higher concentration than their control sample.

We can thus conclude that the spatial distribution of galaxies does not strongly depend on $\Delta m_{12}$. As a consequence, the observed differences in the velocity dispersion profiles should reflect different orbits. This is in agreement with earlier findings by \citet{Girardi2001}, who showed that a peak in the velocity dispersion profile is expected when moderate radial orbits are included, and \citet{Biviano2004} who demonstrated that the central peak can be generated by the presence of late-type galaxies, whereas the profiles of bright ellipticals showed a clear decrease toward the centre.

This theoretical framework, in which differences in the velocity dispersion profiles are connected to different types of orbits, is able to explain in a convincing way the results we found. However, the velocity dispersion profiles of satellite galaxies seem to exclude a link between the orbits and the magnitude gap. In fact, a trend is visible only when including BCGs, whereas it disappears when satellite galaxies alone are analysed. If we compare clusters with the smallest and largest $\Delta m_{12}$ alone (the two most-extreme cases), the former clearly show  a lower velocity dispersion in the centre. But if we include the two intermediate cases the situation changes significantly: the clusters with $1.0 < \Delta m_{12} \le 1.5$  show a clear peak in the central part that excludes any possible trend.

It is worth noticing that clusters with $1.0 < \Delta m_{12} \le 1.5$ are dominated by Abell 85. This cluster has 257 members and the total number of galaxies used in this magnitude-gap bin is 1042. The second most numerous cluster in the bin is FGS31 with 79 members and the mean number of members per clusters (excluding Abell 85) is 35.
The velocity dispersion profile of Abell 85 was already studied in \citet{Aguerri2007} and the authors also found that it peaks in the central part to a value that is larger than that in the external regions. Thus, a possible explanation to the peculiar shape of the velocity dispersion profile for the clusters in the magnitude-gap bin $1.0 < \Delta m_{12} \le 1.5$ is that the subsample is dominated by Abell 85. To test this scenario, we removed this cluster from our sample and we computed again the velocity dispersion profile. The result does not change, meaning that Abell 85 is not the only cluster in this magnitude-gap bin to show this behaviour. We thus have to conclude that the peak in the central part of the velocity dispersion profile is a characteristic of these systems and not caused by a single peculiar case.

The radial velocity dispersion profile was also analysed by \citet{Ribeiro2010}. In particular, they studied a sample of galaxy groups and divided them in two categories: groups with Gaussian velocity distribution (e.g. relaxed) and groups with non-Gaussian velocity distribution (e.g. non relaxed). Non-relaxed clusters were found to have an increasing velocity dispersion profile from the center to R200, while relaxed clusters were found to have a flat velocity dispersion profile in the same radial range. On the other hand, \citet{Cava2017} found the opposite result, the velocity dispersion profiles of irregular clusters declines from the center to R200 more rapidly than that of regular ones. The velocity dispersion profile of our large magnitude-gap systems also show an increasing trend from the center but only out to 0.2*R200. Thus the similarity with the results of \citet{Ribeiro2010} for the irregular clusters is only apparent (compare our Fig.3 with their Fig.2) and it would be erroneous to conclude from this similarity that our large magnitude-gap systems are irregulars. As a matter of fact we showed in Paper VII of this series that the fraction of systems with substructures (i.e. irregular systems) is the same in fossil and non-fossil samples

\section{Conclusions}
\label{conclusions}
We divide a sample of 102 clusters and groups of galaxies into four bins of magnitude gap ($\Delta m_{12}$) in order to study the dependence of the velocity segregation of their galaxies and of their radial velocity dispersion profiles on the $\Delta m_{12}$ parameter. The results we find can be summarised as follows:
\begin{itemize}
\item Velocity segregation only appears in the two brightest bins when it is computed in bins of absolute magnitudes.
\item Velocity segregation only appears in the brightest bin of relative magnitudes.
\item Velocity segregation is limited to central galaxies alone. Satellite galaxies show no segregation independently of their magnitude and of the magnitude gap of their host cluster.
\item The radial profile of the  velocity dispersion shows a trend in the central part with the magnitude gap. In fact, the larger the magnitude gap, the smaller the  velocity dispersion at the centre.
\item The differences in the radial profiles are concentrated within 0.15$R_{200}$. At larger radii, no significant difference is found.
\item The trend disappears if we exclude the BCGs from the computation of the radial profile of the  velocity dispersion. This means that the trend is due to central galaxies alone.
\item A different behaviour appears in the central part of the profiles computed with satellites alone for clusters with $1.0 < \Delta m_{12} \le 1.5$: the central  velocity dispersion is higher in the centre than at large radii, whereas all other systems show an almost monotonic growth.
\end{itemize}

These results show that there is a link between the magnitude gap of the hosting cluster and its central galaxy, but on the other hand the satellite population seems to  show peculiar behaviours not linked to the gap itself. For the BCGs the difference could lie in an earlier formation epoch of the host halo as well as in a more centrally-concentrated halo of the hosting cluster. These scenarios can favor the relaxation of the central galaxy in the centre of the potential well, because they offer a way to accelerate this process: a longer time for dynamical friction to act (in the former scenarios) or a stronger drag due to a larger amount of mass located in the very centre of the cluster halo (latter scenario). On the other hand, the differences in the satellite population could originate from different orbits. We plan to study the orbits of galaxies in FGs in a forthcoming paper of this series.

\begin{small}
{\it Acknowledgements}: The research leading to these results received funding from the European Research Council under the European Union's Seventh Framework Programme (FP7/2007-2013)/ERC grant agreement n. 340519. Moreover, SZ acknowledges financial support from the University of Trieste through the program ``Finanziamento di Ateneo per progetti di ricerca scientifica (FRA2015)'' and from the grant PRIN INAF2014-1.05.01.94.02. 
J. A. L. A. thanks the support from the Spanish Ministerio de Economia y Competitividad (MINECO) through the grant AYA2013-43188-P. 
EMC acknowledges financial support from Padua University through grants DOR1699945, DOR1715817, DOR1885254, and BIRD164402.
ED acknowledges the hospitality of the Institute for Theory and Computation (ITC) at Harvard University, where part of this work has been completed.
\end{small}

\bibliography{bibliografia}{}

\end{document}